\documentclass[12pt]{article}

\usepackage{amsmath,amssymb,graphics,graphicx,epsfig,multirow,color,subfig,amsfonts,amsbsy,mathtools,xargs,tikz,float,rotating}
\usepackage{graphicx,color,url}
\usepackage[colorlinks=true]{hyperref}
\usepackage{epstopdf} 

\newcommand{\SSEA}[2]{#2}

\newcommand{\RWC}[1]{{\color{black}  #1}}
\begin{document}

\title{Optimization of vaccination for  COVID-19 in the midst of a pandemic}

\author{Qi Luo, Ryan Weightman,   Sean T. McQuade, Mateo Diaz,\\
Emmanuel Tr\'{e}lat, William Barbour, Dan Work,\\ Samitha Samaranayake,  Benedetto Piccoli}


\maketitle

\begin{abstract}
    During the Covid-19 pandemic a key role is played by vaccination to combat the virus. There are many possible policies for prioritizing vaccines, and different criteria for optimization: minimize death, time to herd immunity, functioning of the health system. Using an age-structured population compartmental finite-dimensional optimal control model, our results suggest that 
the eldest to youngest vaccination policy is optimal
to minimize deaths. Our model includes the possible infection
of vaccinated populations.
We apply our model to real-life data from the US Census for New Jersey and Florida, which have a significantly different population structure. We also provide various estimates of the number of lives saved by optimizing the vaccine schedule and compared to no vaccination.
\end{abstract}


\section{Introduction}
\subsection{Background}
 Coronavirus disease 2019 (COVID-19) has changed the world in such a massive way that a predictable influx of epidemiological models has been racing to accurately describe the movement of the disease since onset. As the first wave of vaccines hit the general public in early 2021, so did a new wave of models attempting to describe vaccine distribution and the different strategies followed around the world. Vaccination specifically helps develop immunity to the severe acute respiratory syndrome coronavirus 2 (SARS-CoV-2) and possibly limits the spreading of the disease through communities (the restriction of disease spread by vaccine has not been made  clear at this point in time). However, decisions regarding who should be offered COVID-19 vaccines first when supplies are limited remain debatable. 

    \RWC{The institution tasked with minimizing negative outcomes due to disease in the US is the  Center for Disease Control and Prevention (CDC).} The phase allocation plans of COVID-19 vaccines provided by the CDC have addressed the importance of keeping a balance between the prevention of morbidity and mortality and the preservation of societal functioning  \cite{dooling2020advisory}. 
The residents of long-term care facilities are at the highest risk for infection and severe illness from COVID-19, which lists them on the top of vaccination plans to prevent morbidity and mortality.
While less prone to severe infections, healthcare and other front-line essential workers are at higher risk of infection, and their well being is critical for the continuity of essential functions of the society. 
As a result, different social groups are given priority for vaccinations globally. 
For example, most states in the U.S. and Europe prioritize the senior groups, but Indonesia first gives vaccines to essential workers. Speculation about whom to prioritize differs vastly as the general consensus is that healthcare workers and the elderly should come first, but there are many confounding factors. For example, disadvantaged communities can have trouble accessing the vaccine and thus may need to be prioritized as well \cite{subbaraman2020gets}.
A closely related question is how different vaccination plans affect the transmission of a disease over a period of time and the achievement of herd immunity (resistance to the spread of an infectious disease within a population due to large amount of immunity caused by previous infection or vaccination). For example, a leading question is the following: does the vaccination of communities with heightened interaction rates but a lower mortality rate cause less fatalities in the long run? \RWC{In this paper we put a novel spin on an existing method of modeling disease spread. We couple an extended SIR model with an optimization problem in order to show that through such a method, not only can one test possible outcomes, but a model like this can also directly inform policy decisions such as what subset of people to vaccinate first. Our approach couples an age based interaction matrix with public US job records in order to develop vaccination schedules for various states in the United States.}

\subsection{Related works}
Epidemiological models date back to 1932.
The seminal work \cite{kermack1932contributions}
of Kermack and McKendrick subdivided populations into compartments, and provided differential equations driven by infection and recovery rates. The model is also called SEIR
because of the name of compartments: susceptible, exposed, infected and removed.
Many generalizations were proposed, including:
1) considering time-dependent parameters, travel and zoonotic infections \cite{chen2020time,ruktanonchai2020assessing};
2) increasing the number of compartments to capture disease
progression
\cite{giordano2020modelling}; 
3) considering age-structure and spatially distributed populations  
\cite{Britton846,colombo2020well,colombo2020age,Chyba2020a,zhang2020evolving,zhang2020changes}, 4) considering in-host dynamics \cite{bellomo2020multi}.
For the specific case of New Jersey,
a compartmental model was instrumental in  quickly identifying the hospital beds need \cite{allred2020regional} by county, and was cited in a letter from New Jersey Governor Phil Murphy \cite{Murphy2020}. 
Compartmental models are useful for large-scale optimization, a fact we take advantage of in the optimization of vaccine schedule using our model. 
However, they have limitations
in lacking description of in-host dynamics, based only on aggregated data, and
difficulties in representing spatial component of the dynamics \cite{KimQuaini2020}.
Various work discussed the usefulness
of mathematical epidemiological models
\cite{Hyman17,metcalf2020mathematical,vespignani2020modelling}.

From onset of the virus, researchers raced to build models describing how populations could slow the disease using Non-Pharmaceutical Interventions (NPIs).
With uncertainty surrounding vaccine efficacy's duration, the modeling of different possible futures allows us a glimpse at the policy decisions that should be made. 
In \cite{borchering2021modeling} a model is developed to estimate future case counts in the United States by varying projected vaccine efficacy and  NPI policies such as social distancing, mask wearing, and testing. 
In \cite{luo2020managing}, a spacial epidemic model is coupled with an optimization technique 
to resolve safety-and-mobility trade-offs in epidemic response plans.\\
In June 2020, \cite{block2020social} built an agent–based model to predict the effects of social distancing and isolation on transmission of a virus. 
Another agent-based model developed by \cite{patel2021association} is used in order to simulate different vaccination strategies based on projected efficacy of a vaccine in fighting the virus in North Carolina.  
It is important to note that most of these models are considering the minimization of deaths. As seen by the United States Stimulus programs,  \RWC{which are programs developed by the US government in order to jump start the economy through direct payments to citizens, loans to companies and more}\cite{StimProg}, this pandemic has had a huge economic impact in many countries as well. In \cite{shaker2021cost}, rather than using deaths as the cost, a model is developed in order to minimize economic cost to the governments of the United Kingdom, Canada, and the United States.

In late 2020, as vaccinations started to become available, it became clear that this would be the driving intervention to the COVID–19 pandemic.  
In \cite{matrajt2021vaccine}, an age based model was developed before the release of the vaccines 
to predict the 
impact on the virus.  Effectiveness for full vaccine coverage in preventing hospitalization in the age group $65-75$ for example was estimated at $96\%$ for Pfizer-BioNTech, $96\%$ for Moderna, and $84\%$ for Janssen according to \cite{MolineEtal}. 
In \cite{brosh2021bnt162b2}, a study in Israel, it is shown that patients with more comorbidities are much more likely than the average to develop serious infection.
Researchers also looked for optimal vaccination strategies
and estimate the affect the vaccine would have on certain populations.
In fact in \cite{Roghani2021} it is mentioned that the effects of the vaccine on a given population are not observable for quite some time. Therefore models are imperative in predicting and optimizing a best strategy. In \cite{hoertel2021optimizing}, a model was made to examine how long it would take vaccination to be a strong enough intervention to raise all NPIs. 

While evaluating the efficacy of vaccines is important, determining a vaccine schedule is crucial for each governing body during a pandemic. In \cite{bubar2021model}, a compartmental model
is used to address age-based questions about vaccination distribution.  
The seemingly strongest strategy to minimize deaths is the obvious one of vaccinating from the oldest down to the youngest populations.
Similarly, in  \cite{foy2021comparing} a compartmental model is developed in order to answer the question of how to distribute vaccines in India.  One key difference in the model presented in this paper is the stages at which a person is vaccinated. In \cite{foy2021comparing}, the vaccine is given to only susceptible while in our model we allow the vaccine to be distributed to exposed and infected populations as well.

 Previous work in control of pandemic also includes: quantifying the effect of containment measures \cite{Gatto2020},
determining the controllability
using daily data \cite{casella2020can}, considering individual reaction to non-pharmaceutical interventions \cite{lin2020conceptual}, determining best timing
of interventions \cite{gevertz2021novel,perkinsoptimal}, including testing and quarantining
\cite{aronna2021model}.
Moreover, some of the considered interventions were already modeled 
for other viruses such as human papillomavirus \cite{BROWN2011126,saldana2019optimal}.
Finally, some papers focused on the economic cost considering
uncertainty in data \cite{gollier2020pandemic}, cost of lockdown
\cite{acemoglu2020optimal,alvarez2020simple}, hospital and intensive care unit occupancy
\cite{Chyba2020b,Portugal21}.


\subsection{Contribution of the paper}

The focus of this paper is to develop a finite-dimensional model which couples a system of ordinary differential equations with a numerical optimizer to design optimal vaccination strategies \SSEA{to control the spread of COVID-19 by minimizing the mortality amongst a population}{to minimize the mortality amongst a population due to the spread of COVID-19}. The control in question is the \SSEA{}{vaccination strategy} and the objective the minimization of mortality amongst the populations. There were many NPIs developed early during the pandemic to curb the spread of the virus, such as social distancing, contact tracing and testing. While these methods are extremely beneficial, their economic costs rise almost linearly with their frequency of use. Furthermore, if policies are lifted before the disease is all but eradicated, we see an uptick in both cases and deaths soon after. Therefore it seems the most beneficial intervention both in mortality rate and economically will be the vaccine. Not only have the most prevalent vaccines been shown to be very effective, but also a vaccinated person seems to hold a relatively good immunity to sickness and even possibly a lower rate of transmission and infectivity for a much longer time than a government can financially afford to keep the other interventions in policy. 

Four key factors distinguish our approach from others:\\
1) By splitting the population into age ranges, we are able to develop very specific interaction matrices by age group.\\ 
2) Using CDC work data  \cite{cdc}, we are able to split our age groups into essential worker and non-essential worker categories to get a more realistic picture of interactions in the United States.\\
3) Our model allows vaccinated people to become infected (but not seriously sick). This decision was made after it became clear that while a vaccinated person has strong immunity to sickness, it is unclear how strong the vaccine is in preventing transmission of the virus.\\ 
4) Our model is coupled with a complete optimization approach thanks to the Casadi software \cite{Anderson2019}, thus finding the true optimal solution to our model rather than running various simulations with preset schedules. 

This project began with an augmented age-based SEIRV (Susceptible, Exposed, Infected, Removed, Vaccinated) model which uses an interaction matrix to describe interactions between age groups. This decision was made based on the ample data suggesting that mortality rate for the COVID-19 pandemic is (as expected) closely tied to age. \RWC{ With ``work from home orders'' and ``social distancing mandates'' being put in place, and conversely, the essential worker's inability to follow such mandates, to model these interactions, a simple idea is to put the essential workers of each group into their own categories for which interactions are less affected by mandates.}  The time horizon is chosen to be $180$ days as this seems to be a realistic amount of time to vaccinate the majority of the eligible population (with the optimistic point of view that the entire population would like to be vaccinated). This point of view has been proven in many countries to be very optimistic as has been made clear by vaccine hesitancy and outright opposition in many parts of the world (including in the United States). As a booster shot is being developed and approved at the government level, one sees that human consequences are leading to a lag in herd immunity.  Our interaction matrices are developed using data from \cite{prem2017projecting}. We use age-based death rates developed from \cite{cdc} in order to estimate the expected number of deaths for a given vaccination schedule.\\
We then adjust the model slightly to become an  augmented age-work-based SEIRV model. The big question amongst policy makers seems to be whether it is optimal to prioritize essential workers or to prioritize the oldest populations first.  We capture the competition between these two ideas by breaking our model up using labor force statistics from the Bureau of Labor \cite{Laborstat}. We split all working age populations into categories ``essential'' and ``non-essential'' in order to see if there is a difference in optimal policy if the interactions of essential workers are scaled up. 

We then further develop the model by removing the vaccinated compartment and instead including a vaccinated compartment for each of the usual compartments: \SSEA{Susceptible, Exposed, Infected, Removed, }{} Susceptible Vaccinated, Exposed Vaccinated, Infected Vaccinated and Removed Vaccinated. This decision was made in order to make the model \SSEA{closer}{better} fit the physical properties of the population; a vaccinated person has a high enough probability to become infectious and spread the virus that we felt it crucial to include such dynamics.  With this change, the user can capture the real life implications of  an unknowing exposed or infected individual (or knowing if there is a perceived health benefit)  getting vaccinated. This new model also allows a user the ability to capture an elusive interaction: vaccinated infected-vaccinated susceptible. While there is not much literature about the transference of virus amongst vaccinated susceptible populations for COVID-19, it is more clear than ever that there are two driving forces to a virus: spreadiblity and mortality. If a virus is spread with ease throughout the vaccinated compartments, while those in the vaccinated compartments are assumed safe from serious illness, they are not safe from being hosts through which the virus can travel to a more vulnerable host, thus causing serious illness. Our model opens the door to tests which could show how these interactions between the vulnerable and the vaccinated spreader affect the population as a whole. 

Once our model is built, we tune the parameters to state specific data (in this paper we show results from both New Jersey and Florida) in order to test vaccination schedules. We then \SSEA{settle}{formulate} an optimal control problem, consisting of minimizing the total number of deaths over the time horizon that we numerically solve by a direct transcription using \cite{Anderson2019}, an open-source tool for nonlinear optimization and algorithmic differentiation. Our numerical simulations produce an optimal vaccination schedule for the given parameters and data set. 
As will be mentioned later, it is uncertain what exactly \SSEA{would be considered}{constitutes} an ``essential-worker''  in each state, a value as elusive as the initial replication rate and the scaling of the interaction matrix. Because of this, we test several scenarios with various \SSEA{values of all three of these}{choices as a sensitivity analysis} in order to paint a broader picture. 

The most striking yet \SSEA{relieving}{comforting} result is that in every case, the optimal solution seems to be the one chosen by most policy makers during this pandemic; to vaccinate the oldest population first and then work down the list, vaccinating based on age while always vaccinating the essential worker population first. While this policy seems to be robust, such a choice is not always easy, as seen by countries who chose to vaccinate essential workers first at the onset of vaccine production. The tool that we have designed could be useful in not only the vaccination distribution for this pandemic, but also could be used as a blueprint for future vaccination schedule optimization questions.  

\section{Discussion on Agent Based and Compartmental Models}

There are generally two types of epidemiological models used to examine the trends and fluctuations of populations due to a disease; agent-based models (ABM) and compartmental models (CM). These models differ in a few ways, starting with how they organize a population: a CM focuses on capturing the collective behavior of a population or sub-population as a whole. For example in a standard SIR CM we have three compartments: Susceptible, Infected and Removed. We define differential equations which govern the movements of the population amongst these three compartments. At each time step, the entire population is defined as the sum of the three categories with each member of the populations' only defining characteristic being the compartment that they are in. 
CMs were extensively used to model
infectious diseases, including
pertussis~\cite{rohani2010contact, de2012estimation}, measles~\cite{bokler1993chaos}, Ebola~\cite{lekone2006statistical, mamo2015mathematical},  HIV~\cite{ozalp2011fractional, demirci2011fractional}, tuberculosis~\cite{side2016global, bowong2009mathematical}, cholera~\cite{crooks2014agent} and others. CM models were successfully
used also for COVID-19, for instance in
\cite{zhou2020clinical, Branas2020, Ferguson2020, Giordano2020, Gatto202004978}.

In contrast, an ABM, as the name may suggest, focuses on the behavior of \SSEA{an}{each} individual in the system. Each individual in this case would be considered an ``agent'' and the state of an agent is governed by probabilities based on interactions with other agents. One immediately sees both the merits and the restrictions of an ABM. In considering each agent and all of their defining characteristics, we are able to gain a very detailed view of the progression of a disease all the way down to knowing a few defining characteristics about each and every agent. However, the restriction here is that in order to get such a detailed view, one must input many details about the system. This type of model works best for a population which the researcher is able to describe using many choices and consequences made by each agent and the probabilities which govern these choices on a large scale. As such, an ABM requires large amounts of precise data in order to paint an accurate picture of real world dynamics. This brings up concerns about robustness of a model both in accuracy of inputs and transfer-ability to different data sets.  

This highlights another key difference between the two model types, namely the use of probability. An ABM is at essence a stochastic model using probabilities to govern behavior. Therefore, when using an ABM one must complete multiple simulations and take the average of the solutions in order to ensure an outlier solution has not been found. For example, in \cite{hoertel2021optimizing}, an ABM is used to model vaccine strategy strength, testing whether the vaccines would be robust enough to allow for the removal of non-pharmaceutical interventions in the country of France. The model is run 200 times and results are averaged in order to paint the most probable picture. An ABM, once built, has the capacity to adjust for micro-scale policy changes in a population such as social distancing regulations or mask wearing regulations in an efficient manner. In \cite{kerr2020covasim}, they highlight the ease at which their ABM can be adapted for policy change. However, requiring each member of a population to have such complex characteristics requires the model to remember the state of each individual at each time step and can result in a very costly model. This is where the CM becomes more efficient. While a CM is less easily adapted to new policy, it can scale to any population with much less cost than an ABM. This is because one must not worry about each individual in the population. For example, in \cite{hoertel2021optimizing}, the ABM being used only considers $500,000$ agents despite the population of France being around 67 million people at the time. This need for scaling and then re-scaling often still allows for an accurate model given realistic parameters, but highlights the computational power needed to run an ABM at full capacity. 

An advantage here for the CM is that the substitution of a different data set, for example using the same model for two different states in the United States, results in a very minor difference in cost. This is a reason why in this paper we will describe the use of a CM to model vaccine distribution in the united states. While different states have vastly different populations and age distributions, the same optimization problem can be solved on them using the same CM with predictable difference \SSEA{is}{in} cost. In this paper we use data from the 50 states, treating each state's vaccination schedule as its own optimization problem.  Therefore, the fluctuating populations suggests that a CM is better suited here and thus is what we employ.

\section{An age-structured compartmental model}
Conventional epidemiological models assume the homogeneity of different social groups as well as the effectiveness of vaccines. 
To personalize the vaccination phase allocation with a balance of mortality prevention and societal function preservation, we first build an age-and-work structured model. The entire population is divided into seven age groups, and those groups of working age are further divided into essential and non-essential workers. 
The dynamics of the system are described by an age-and-work structured susceptible-exposed-infected-removed (AW-SEIR) model.
Each susceptible individual undertakes the risk of interacting with an infected person, becoming exposed, and after a latent exposed period ends up being recovered or deceased. 

Essential workers are at higher risk of infection due to extended exposure to a larger population of possibly infected people. They could also not be receiving adequate access to personal protective equipment (PPE) in their workplaces early on.  
The senior age groups are at higher risk of morbidity or mortality at the final stage. 
Other groups may also have different social activity levels compared to the pre-pandemic situation. 
Modeling interactions (i.e., social contacts \cite{prem2017projecting}) between different age-and-work groups is certainly a central task. 
The intensity of social contacts within each group and between groups is primarily dependent on various social-distancing measures and behavior changes. 
The locations of those contacts are categorized as work, home, school, and others \cite{prem2017projecting}.
This compartmental model is able to describe a variety of policies such as school closure, enforced work-from-home for non-essential workers, or social-distancing in workplaces all through the interaction matrices defined later. 

The social contacts between different age or work groups could give rise to non-trivial phase allocation plans for vaccine distribution during a pandemic. For instance it is clear that a disease with extremely low morbidity, or less age-based morbidity would require those with more interactions to be vaccinated early. However, a higher age-based morbidity would require the more susceptible ages be vaccinated first. The goal of models such as these is to find a precise plan which takes all factors into account.  

Considering the primary objective of preventing morbidity and mortality, one would naturally assume that the senior-age groups should be given vaccines before others. 
Nevertheless, due to the slow vaccination process, a considerable proportion of essential workers being exposed to the disease may become super spreaders. 
Those super spreaders may bring the virus home and to workplaces so that the infected population grows exponentially because of the failure of adequate protection. 

Even if the mortality rate of those age groups is smaller than the mortality rate of the seniors, the larger basis of population with severe symptoms may exhaust limited healthcare resources and lead to unexpected consequences.
This consideration is valid especially when the basic reproduction number $R_0$ is large. 
The interaction between those vulnerable societal groups and active social groups plays a prominent role in determining the weights on allocating the limited supply of vaccines. 

The phase allocation of vaccines is formulated as an \emph{optimal control problem} based on this AW-SEIR model. 
In each time step, the decision-maker observes the susceptible population of each societal group and determines the number of vaccine doses that should be allocated for each group.
For each societal group, the vaccination decision made in the current period will not only affect the prevented number of COVID-19 cases in this group but have a spillover effect on other groups through social contacts; vaccination of any person reduces the chances of that person to become infected, but also reduces the chances of that person infecting others.
Therefore, the controller must optimize the vaccination plans for all groups synchronously with the objective of reducing the total mortality in the population before reaching herd immunity.

The age structured epidemic model with vaccination divides the total population into 11 groups in  Table \ref{tab:AgeGroups}.

\begin{table}[ht] 
 \begin{tabular}{|p{1in}|p{3.5in}|} 
 \hline
 Name & Description  \\ [0.5ex] 
 
 \hline
 Group 1 & Age 0-4 population \\
 \hline
 Group 2 & Age 5-14 population \\
 \hline
 Group 3 & Age 15-19 population with no job or non-essential  \\
 \hline
 Group 4 & Age 20-39 population with no job or non-essential  \\
 \hline
 Group 5 & Age 40-59 population with no job or non-essential  \\
 \hline
 Group 6 & Age 60-69 population with no job or non-essential \\
 \hline
 Group 7 &Age 70+ population   \\
 \hline
 Group 8 & 	Age 15-19 population who are essential workers\\
 \hline
 Group 9 & Age 20-39 population who are essential workers \\
 \hline
 Group 10 & Age 40-59 population who are essential workers \\
 \hline
 Group 11 & Age 60-69 population who are essential workers\\ [1ex] 
 \hline
\end{tabular}
\caption{Groups by Age}\label{tab:AgeGroups}
\end{table}

We use the subscript j to indicate a social group. we then  indicate the total population using the variable N, thus  $N_j$ is the number of people in age group $j \in \{1, \dots, 11 \}$. Thus we have:

	$S_j$ the number of susceptible;
	
	$E_j$ the number of latently infected or exposed;
	
	$I_j$ the number of infectious people not isolated;
	
	$R_j$ the number of removed;

	$V_j$ the number of vaccinated.

The dynamics of the age structured SEIR model is given by:
\begin{eqnarray}\label{eq:SEIRV1}
\begin{aligned}
\dot{S_j}&=  - u  \frac{S_j\, \displaystyle \sum_{k=1}^{11} l_{kj}\ I_k}{\displaystyle \sum_{k=1}^{11} l_{kj}\  N_k} - w_j\\
\dot{E_j}&=  u\  \frac{S_j\,\displaystyle \sum_{k=1}^{11} l_{kj}\ I_k}{\displaystyle \sum_{k=1}^{11} l_{kj}\ N_k} -\delta E_j\\
\dot{I_j}&=\delta E_j -\gamma {I_j}\\
\dot{R_j}&= \gamma {I_j}\\
\dot{V_j}&= w_j
\end{aligned}
\end{eqnarray}

where $u = R_0$  $\gamma$ reflects the lockdown measures (and could be time-dependent); $L=(l_{k j})$ is the interaction matrix among age groups during pandemic. In this matrix, the maximal eigenvalue is given by the basic reproduction number $R_0$.  $w_j$ is the number of vaccinated people in age group $j$ per day, so that $\sum_1^{11}w^j$  is the number of vaccines available per day (and could be time-dependent); $\delta=1/D_E$ with $D_E$ being the latent period in days; similarly, $\gamma=1/D_I$ with $D_I$ denoting the infectious period in days. Now it is also important to choose epidemiological constants which are \SSEA{accurate}{specific} to the disease in question. 
These parameters are estimated with CDC sources in Table \ref{tab:VarTable}.

\begin{table}[ht] \begin{center}
 \begin{tabular}{||c c c c||} 
 \hline
 Name & Description & Estimate  & Units  \\ [0.5ex] 
 
 \hline
 
$R_0$  & Rate of infection & 1.0-1.2 & – \\
 \hline
 $D_I$  & Infectious period & 5-14 & days \\
 \hline
  $D_E$  & Latent period & 4-7 & days \\ [1ex] 
 \hline
\end{tabular}
\caption{Description of Variables}\label{tab:VarTable}
\end{center}
\end{table}

\begin{figure}
\begin{center}
\includegraphics[angle=0,
width=5.0cm, height=7cm,
trim={0cm 0cm 0cm 0cm},clip]{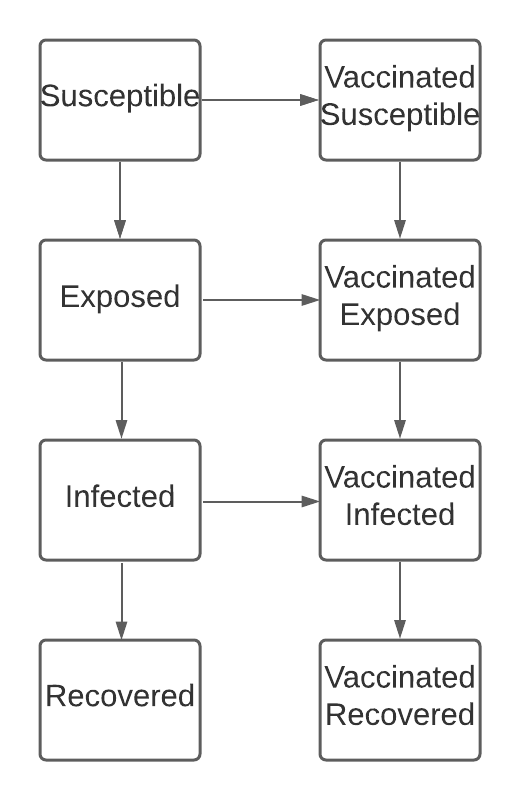}
\caption{All possible paths through which populations may flow into other populations.} \label{fig:ModelDiag}
\end{center}
\end{figure}

After a few weeks of vaccinating, the notion began to spread that perhaps the vaccine not only keeps a person from contracting the disease, but also from developing symptoms once infected and spreading disease if infection is contracted. In \cite{mallapaty2021can}, a driving question is whether the COVID-19 vaccine can stop one from being a spreader of the disease after contracting the infection.  In order to account for modeling these possibilities, we further the complexity of the model by incorporating an expansion of the vaccinated compartment; rather than including V, we add Susceptible Vaccinated (Sv), Exposed Vaccinated (Ev), Infected Vaccinated (Iv) and Removed Vaccinated (Rv) in which we are able to change the parameters which govern the ability of the vaccinated populations to transmit the virus, see Figure \ref{fig:ModelDiag}.

So in the end, our model \SSEA{reads like equation array}{is described by the system of equations} \eqref{eq:SEIRV2}.  

\begin{eqnarray}
\label{eq:SEIRV2}
\begin{aligned}
\dot{S_j}&= - u  \frac{S_j\, \displaystyle \sum_{k=1}^{11} l_{kj}\ I_k+0.3*l_{kj}\ Iv_k}{\displaystyle \sum_{k=1}^{11} l_{kj}\  N_k} - \frac{w_j\ S_j}{S_j+E_j+I_j}\\
\dot{E_j}&=  u\  \frac{S_j\, \displaystyle \sum_{k=1}^{11} l_{kj}\ I_k+0.3*l_{kj}\ Iv_k}{\displaystyle \sum_{k=1}^{11} l_{kj}\  N_k} -\delta E_j-\frac{w_j\ E_j}{S_j+E_j+I_j}\\
\dot{I_j}&=\delta E_j -\gamma {I_j}-\frac{w_j\ I_j}{S_j+E_j+I_j}\\
\dot{R_j}&= \gamma {I_j}\\
\dot{Sv_j}&= - u  \frac{Sv_j\,\displaystyle \sum_{k=1}^{11} 0.3*l_{kj}\ I_k+0.09*l_{kj}\ I_k}{\displaystyle \sum_{k=1}^{11} l_{kj}\  N_k}+\frac{w_j\ S_j}{S_j+E_j+I_j} \\
\dot{Ev_j}&=  u\  \frac{Sv_j\, \displaystyle \sum_{k=1}^{11} 0.3*l_{kj}\ I_k+0.09*l_{kj}\ I_k}{\displaystyle \sum_{k=1}^{11} l_{kj}\  N_k} -\delta Ev_j+\frac{w_j\ E_j}{S_j+E_j+I_j}\\
\dot{Iv_j}&=\delta Ev_j -\gamma {Iv_j}+\frac{w_j\ I_j}{S_j+E_j+I_j}\\
\dot{Rv_j}&= \gamma {Iv_j}
\end{aligned}
\end{eqnarray}
\RWC{In the final model our compartments include the usual SEIR with four additional compartments representing vaccinated populations. we have the interaction matrix $l_{kj}$ governing interactions, $w_j$ governing total vaccine allocation in a day, $N_k$ being the total population, $\delta$ being the virus mean latent period and $\gamma$ being the mean infectious period.  } There is reduced probability of an infected but vaccinated person to transmit the disease. We chose to model this by a factor of $0.3$. \RWC{It is important to note that this choice is an estimate as there is not much reliable data in the literature about how the vaccine effects transmission. However, \cite{singanayagam2021community} claims that vaccination was found to be effective in diminishing transmission inside of the household to forty percent. Therefore thirty percent seems a reliable estimate for transmission outside of the household with other non-pharmaceutical interventions at play. } As such, a vaccinated-vaccinated interaction has an estimated reduced chance of transmission to $0.09$.  This new model accounts also for the ability to vaccinate other compartments in addition to $S$ (Susceptible): terms such as $\frac{w_j\ S_j}{S_j+E_j+I_j}$ show the administration of the vaccines. This is because with the influx of vaccination there was also a diminishing of testing. The result is that an ``exposed'' or ``infected'' but asymptomatic person could get the vaccine. 


\section{Census Data and Interaction Matrix}
In order to test our model, we use collected data to develop initial conditions for S, E, I and R, interaction matrices, death rates, basic reproduction number and infection rates and recovery rates for our age groups. The majority of the data used comes from the United States census. 

We first take the raw population information from each state in the united states and use the Age and Sex tables provided to break the population in each state down by age group. From here we fit the populations into the age groups defined below for our model. We then define the number of susceptibles to be the number of people in each age group after removing the exposed infected and removed; the number of exposed is set equal to a certain percent of the infected from the first 5 days of the model (Dec 16,17,18,19,20); the number of infected is set to be the number of cases from the first 5 days of the model minus exposed assigned proportionally based on the number of people in each age group; the number of removed is set to be all cases from January up to and including cases on Dec 15 assigned proportionally based on the number of people in each age group.

As the definition of the term ``essential worker'' is a bit ambiguous, we must also justify our percentages there. We begin with  labor force statistics from the Bureau of labor. We then decide which occupations count as essential. This decision was based on descriptions of essential workers from \cite{suzannehultin_2021}. We then sum the total workers in each age group and reassign  people to the essential worker category based on our age groups (the age groups provided by the Bureau of Labor did not perfectly match our age groups). Lastly we use the total population from the CDC age groups to calculate an “essential worker rate” per age group.

The basic reproduction number (to be used to scale $A$) was found per state at rt.live \cite{rt}, a live Covid tracking website which recalculates and reports daily replication numbers. 

The total vaccine availability per day $w$ was estimated by CDC sources. \RWC{We chose this number to be $10,000$ as this is both a rational assumption for a state in the US and a good estimate for herd immunity by sixth months, which was a goal of policy makers.} Of course this is a simplification of the process as (what we are now seeing live) the number of vaccines administered is reduced after the more eager portion of the population has been successfully vaccinated.

We construct the interaction matrix using $A$ and $B$: $\alpha$ is chosen so that the dominant eigenvalue (call this $\lambda_{max}$) of the matrix $\alpha (A-B) + \beta B$. 
 The eleven age groups are$\{1 ,2 , 3 , 3E , 4 ,$ $ 4E , 5 , 5E , 6 , 6E , 7 \}$. We use a combination of $A$ the interaction matrix, and $B$ the working interaction matrix for groups 1 through 7 to build a new interaction matrix which describes the interactions of nonessential and essential age groups. 

Matrices $A$ and $B$ are shown below. $A$ gives the interaction coefficients which represent how often members of different age groups are exposed to each other. $B$ gives the interactions only due to the working environment. Therefore, we subtract the interaction coefficient given in $B$ from that in $A$ to describe the interaction coefficient of a Non-essential worker in an age group, because they are assumed to not have working interactions.

\begin{align*}
    A= \begin{bmatrix}
    2.5982 & 0.8003 & 0.3160  & 0.7934 & 0.3557 & 0.1548 & 0.0564 \\
0.6473 & 4.1960 & 0.6603  & 0.5901 & 0.4665 & 0.1238 & 0.0515 \\
0.1737 & 1.7500 & 11.1061 & 0.9782 & 0.7263 & 0.0815 & 0.0273 \\
0.5504 & 0.5906 & 1.2004  & 1.8813 & 0.9165 & 0.1370 & 0.0397 \\
0.3894 & 0.7848 & 1.3139  & 1.1414 & 1.3347 & 0.2260 & 0.0692 \\
0.3610 & 0.3918 & 0.3738  & 0.5248 & 0.5140 & 0.7072 & 0.1469 \\
0.1588 & 0.3367 & 0.3406  & 0.2286 & 0.3673 & 0.3392 & 0.3868
    \end{bmatrix}
\end{align*}

\begin{align*}
    B=\begin{bmatrix}
    0      & 0      & 0      & 0      & 0      & 0.0000 & 0.0000 \\
0      & 0.0195 & 0.0094 & 0.0116 & 0.0115 & 0.0000 & 0.0000 \\
0      & 0.0168 & 0.6441 & 0.3590 & 0.1893 & 0.0067 & 0.0000 \\
0      & 0.0272 & 0.3060 & 0.8135 & 0.5203 & 0.0218 & 0.0000 \\
0      & 0.0361 & 0.2103 & 0.6465 & 0.6465 & 0.0234 & 0.0000 \\
0.0000 & 0.0079 & 0.0083 & 0.0585 & 0.0692 & 0.0035 & 0.0000 \\
0.0000 & 0.0000 & 0.0000 & 0.0000 & 0.0000 & 0.0000 & 0.0000
    \end{bmatrix}
\end{align*}
We use $A$ and $B$ to build an 11$\times$11 matrix for age groups 
\[
\{1 ,2 , 3 , 3E , 4 , 4E , 5 , 5E , 6 , 6E , 7 \},
\]
where $3$ is the third youngest age group (age 15 to 19) excluding Essential workers (``Non-Essential Workers''), and $3E$ is the group of Essential workers in this age group.
We define the Nonessential group to Nonessential group interactions as $\alpha (A- B) $. This is an 11$\times$11 matrix with zeros on the rows and columns with essential workers ($\{3E ,  4E , 5E , 6E \}$) and entries from $\alpha (A- B) $ on the other rows and columns. The Nonessential to Essential interactions have entries from $\alpha (A- B) + \beta B$ on 
For the total interactions matrix, we have:
\[
\begin{aligned}
I &= [C_1 (A- B)] + [C_1 (A- B) + B] + [C_1 (A- B) + B] + [C_1 (A- B) + B],
\end{aligned}
\]
where $C_1$ is a scaling factor to add weight to the nonworking interactions.

\section{Optimization of vaccination schedule}
We consider an optimal control problem for the system \eqref{eq:SEIRV2},
with cost given by the total number of deaths over a time horizon of 180 days. The model was tuned with data from New Jersey and Florida on January 1st 2021.
The number of available vaccines was set so in 180 days all individuals above 15 years of age could be vaccinated.

In order to solve an optimal control problem numerically, it is usual and well known to distinguish between \emph{direct} and \emph{indirect} methods. 

Roughly speaking, direct methods (or direct transcription) consist in fully discretizing the optimal control problem under consideration, before applying an optimization routine. Discretizing means: choose a discretization scheme for the differential equations (like Euler, Runge-Kutta); choose a discretization for the integral cost criterion (like rectangle, Simpson). One can also approach the solutions globally in time by pseudo-spectral methods, by collocation. There are infinitely many choices. In all cases, the full discretization yields a classical nonlinear optimization problem under constraints, in finite dimension (this dimension being larger as the discretization is finer). 
This high-dimensional nonlinear optimization problem can then be solved thanks to numerical optimization: gradient methods, penalization, Lagrange methods (Karusch-Kuhn-Tucker). There also, one has infinitely many choices to design a numerical scheme. 
We note that, in direct methods, we first discretize, then optimize (or dualize). 

In the indirect approach, this is exactly the other way round: we first apply a first-order necessary condition for optimality to the optimal control problem: we apply the \emph{Pontryagin maximum principle} (see \cite{BressanPiccoli,LeeMarkus,Pontryagin,Trelat_book}), which leads to a \emph{shooting problem} that can then be solved, numerically, thanks to a Newton method (see \cite{BCT2007} for well-posedness of the shooting method, in relationship with conjugate point theory). In other words, in the indirect approach, we first optimize (or dualize) and then discretize. 

We refer the reader to \cite{trelat_JOTA2012,Trelat_book} for a survey on the numerical methods in optimal control and on the pros and cons of direct vs indirect approaches. Also, note that many existing numerical methods are neither direct nor indirect but are rather of a hybrid nature. The choice of such or such method depends on the context and on the demanding issues.

Here, we choose the direct transcription approach because the optimal control problem that we consider involves some state constraints that would be difficult to handle in the indirect approach (indeed, using the Pontryagin maximum principle when there are state constraints is much more involved). Moreover, direct methods are much softer insofar they allow to easily perform modifications of the model. 

We choose to discretize the control system with the implicit RK2 scheme and the cost functional with the trapezoidal rule, on a regular subdivision of the time interval (we take: one step = one day). Writing the control system as $\dot x(t) = f(x(t),u(t))$ and the cost functional as $C(u) = \int_0^T f^0(x(t),u(t))\, dt$, this means:
\begin{align*}
& x_{k+1} = x_k + \frac{h}{2} \left( f(x_k,u_k) + f(x_{k+1},u_{k+1}) \right), \qquad k=0,\ldots,N-1 \\
& \min \frac{h}{2} \sum_{i=0}^N \left( f^0(x_k,u_k) + f^0(x_{k+1},u_{k+1}) \right)
\end{align*}
where $h$ is the step of the (assumed) regular subdivision of the time frame $[0,T]$.

As an optimization routine to solve the resulting (finite-dimensional) optimization problem, we use the interior-point optimization routine \texttt{IpOpt}~\cite{Waechter2006}, combined with the modeling language \texttt{AMPL} which performs automatic differentiation. It is by now well known that the use of automatic differentiation is a real plus in solving efficiently nonacademic, nonobvious optimal control problems that cannot be solved so efficiently without this powerful tool.

To initialize the optimization, we just take constant (discretized) controls and states, those constants, for the states, corresponding to their initial conditions. This is a very rough initialization but it happens to be enough for our needs. The execution on a standard desktop machine is almost instantaneous. 


One of the goals is to compare the policy to vaccinate based on age (oldest to youngest)
opposed to work status
(essential workers first) structure. Note that in order to test the elasticity of our model, we test varying the initial replication rate $R0$ $(1.0, 1.1, 1.2)$, the percent of workers (PE) deemed ``essential''($24\%$, $34\%$, $44\%$) and the infection rate $\beta$ ($0.25$,$0.5$, $0.75$) of COVID-19. 

\begin{figure}
\begin{center}
\includegraphics[angle=0,
width=13.0cm, height=7cm,
trim={0cm 0cm 0cm 0cm},clip]{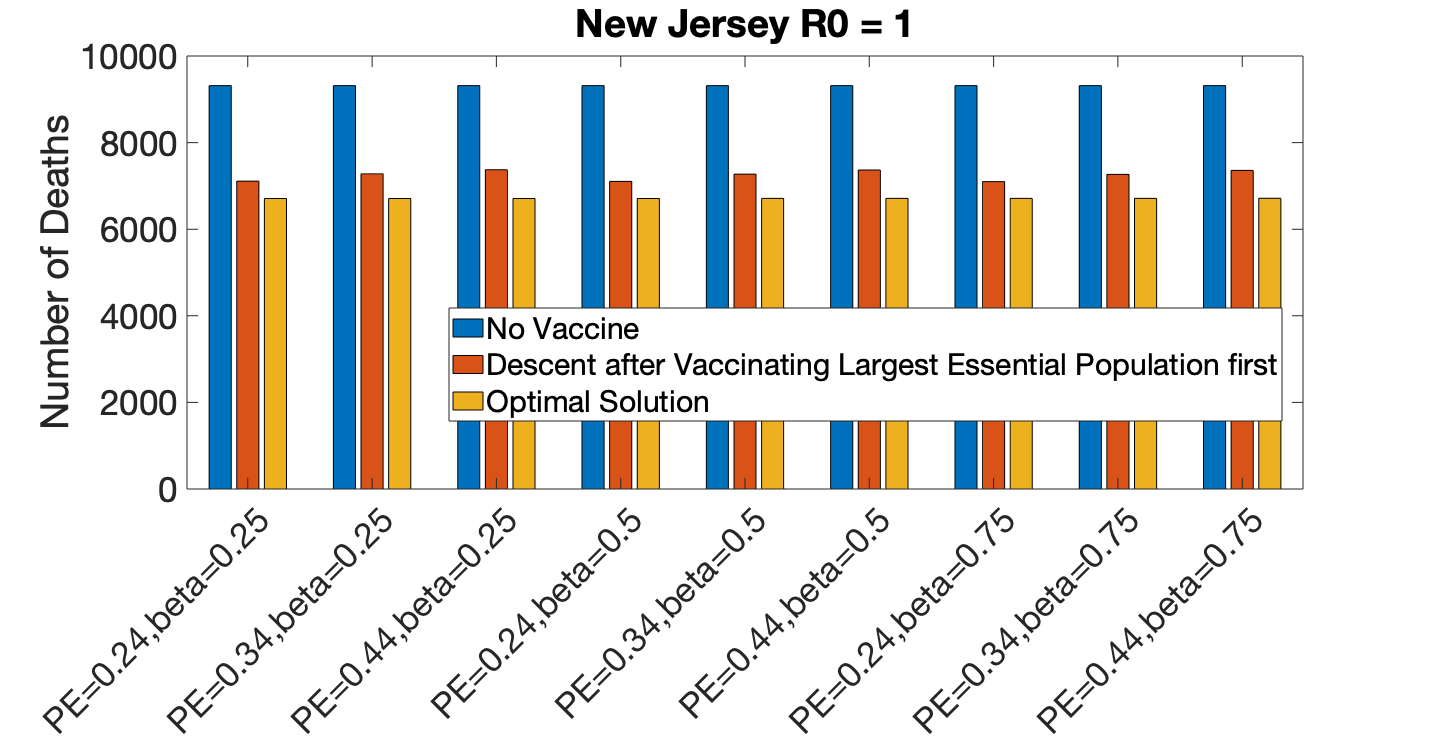}
\caption{Sample tests for New Jersey with a choice of $R0 = 1.0$ (mildest case) while varying percent of essential worker and beta. } \label{fig:NJR01.0}
\end{center}
\end{figure}

    \begin{sidewaysfigure}
\vspace{30mm}
\hspace{0mm}
\includegraphics[
trim={0cm 0cm 0cm 0cm},clip]{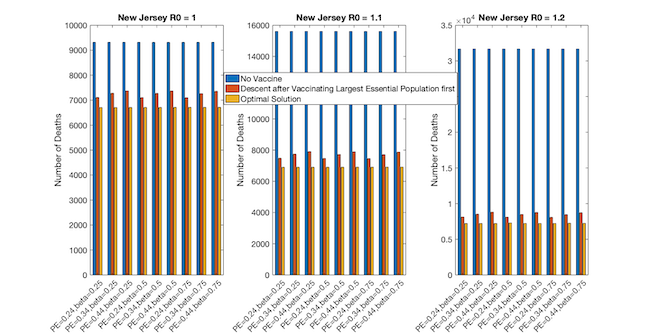}
\caption{Results using New Jersey data-set plotted by initial replication rate.} \label{fig:NJTests}

\end{sidewaysfigure}

    \begin{sidewaysfigure}
\vspace{30mm}
\hspace{0mm}
\includegraphics[angle=180,
trim={0cm 0cm 0cm 0cm},clip]{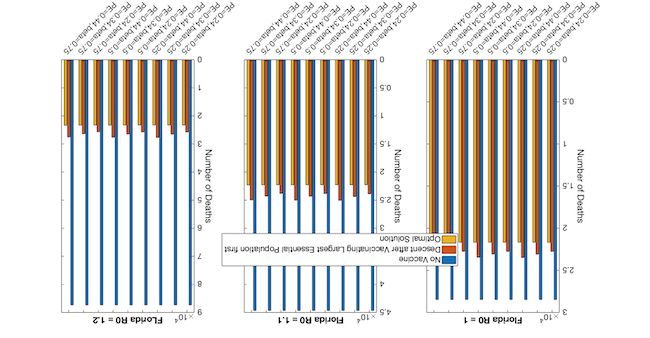}
\caption{Results using Florida data-set plotted by initial replication rate.} \label{fig:FLTests}

\end{sidewaysfigure}

    One reassuring result which can be seen clearly in Figure \ref{fig:NJR01.0} is that any amount of vaccination greatly diminishes the number of deaths regardless of who is being vaccinated, even in the case of a very mild disease. A case where $R0=1.0$ is a very benign disease, yet we still see that a vaccine will save about $2000$ lives even in the non-optimal case. 
    

    However, if $R0=1.2$, we see in Figure \ref{fig:NJTests}  that the number of lives saved jumps to $23,500$ immediately regardless of other parameters. Therefore, it is very clear that the vaccine, regardless of schedule, is extremely effective in minimizing mortality across all categories. A small rise in $R0$ results in almost an exponential rise in deaths in an unvaccinated population, but we see that any vaccination plan diminishes this jump significantly. This can be clearly seen in  Table \ref{tab:DeathsTable} where we keep the number of essential workers and beta constant and vary the $R0$. Further, we see that while varying the PE and \SSEA{the}{} $\beta$ changes the number of deaths (\SSEA{raising}{increasing} both results in a rise in deaths), the driving factor behind the consequences of COVID-19 is indisputably the infectivity of the virus. This can be seen very strongly in Figure \ref{fig:FLTests} where we see that raising the percent of essential workers raises the number of deaths by less than one hundred in some cases and not much more in others, compared to the  few hundred lost when raising the replication rate of the virus even in the optimal vaccination case.  
    
      \begin{table}[ht] \caption{Deaths Projected with Varying $R0$}\label{tab:DeathsTable}
    \begin{tabular}{|p{1in}|p{.5in}|p{1.5in}|p{1.5in}|} 
 \hline
 State & $R0$ & Projected Deaths With No Vaccine & Projected Deaths With  Vaccine  \\ [0.5ex] 
 
 \hline
 New Jersey & $1.0$ & 9316 & 6710\\
 \hline
 New Jersey & $1.1$ & 15609 & 6906\\
 \hline
 New Jersey & $1.2$ & 31681 & 7289\\
 \hline
 Florida & $1.0$ & 28467 & 21678 \\
 \hline
 Florida & $1.1$ & 44657 & 22287 \\
 \hline
 Florida & $1.2$ & 87349 & 23298 \\
 \hline

\end{tabular}
\end{table}
\vskip 5mm
  
\begin{figure}
\begin{center}
\includegraphics[angle=0,
width=14.0cm, height=14cm,
trim={0cm 0cm 0cm 0cm},clip]{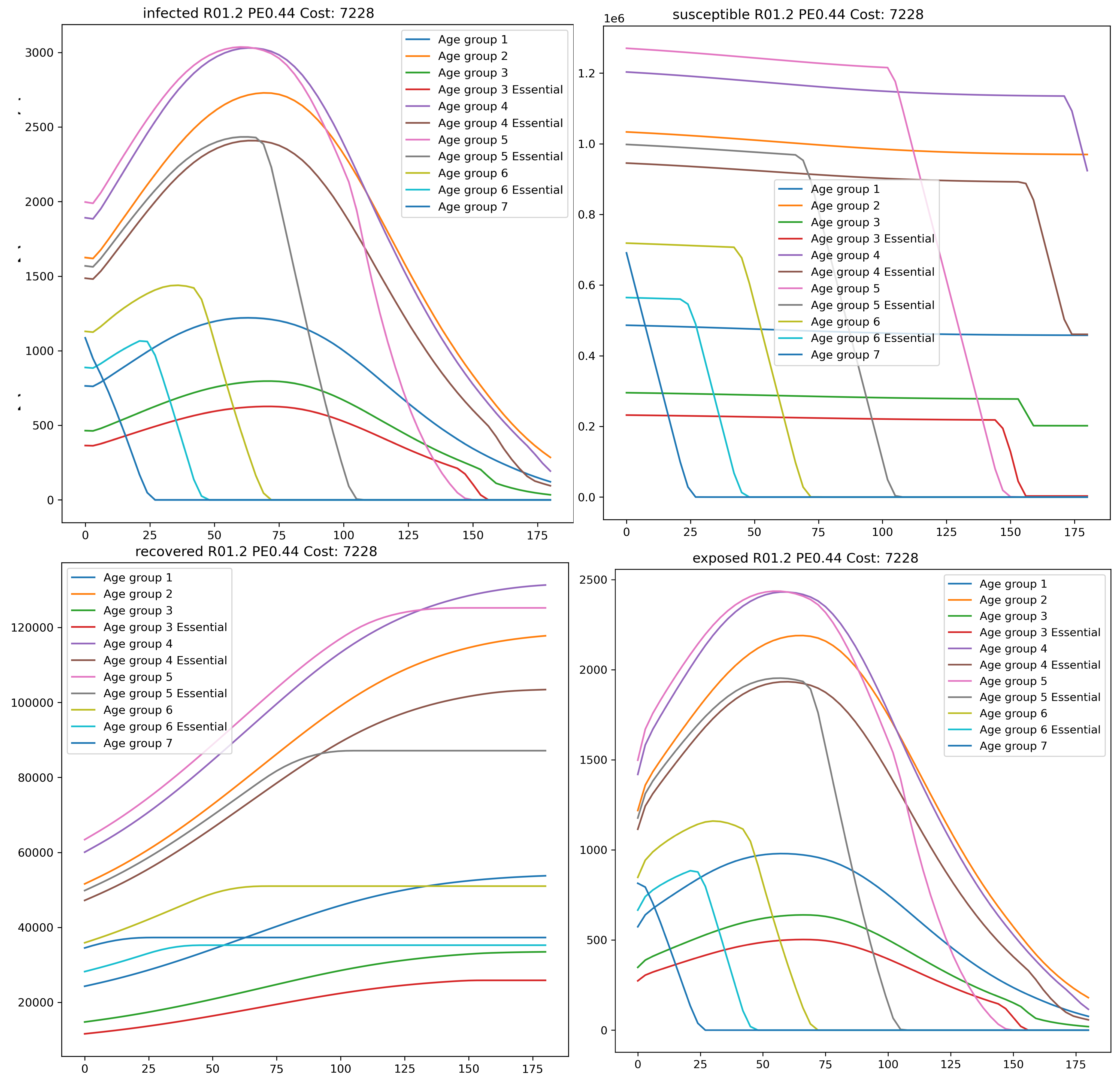}
\caption{\RWC{Population dynamics for the unvaccinated compartments: Susceptible, Exposed, Infected, and Recovered.} } \label{fig:pops}
\end{center}
\end{figure}

    Note that in the dynamics of our program such as the exposed population in Figure \ref{fig:pops} we see that the program terminates at $180$ days. This decision to choose a time-span of 180 days in our program is due to the simple fact that the most susceptible populations are vaccinated by then and thus the virus is virtually non-existent beyond this point. This can clearly be seen in the infected population in Figure \ref{fig:pops} where the infected populations are all but diminished and only those with extremely low death rates are still infected. If this policy happened in reality, while the virus may still permeate on a biological level, the sickness and death rates would be so low that in the eye of the public it would be non-existent. 

\begin{figure}
\begin{center}
\includegraphics[angle=0,
width=9.0cm, height=7cm,clip]{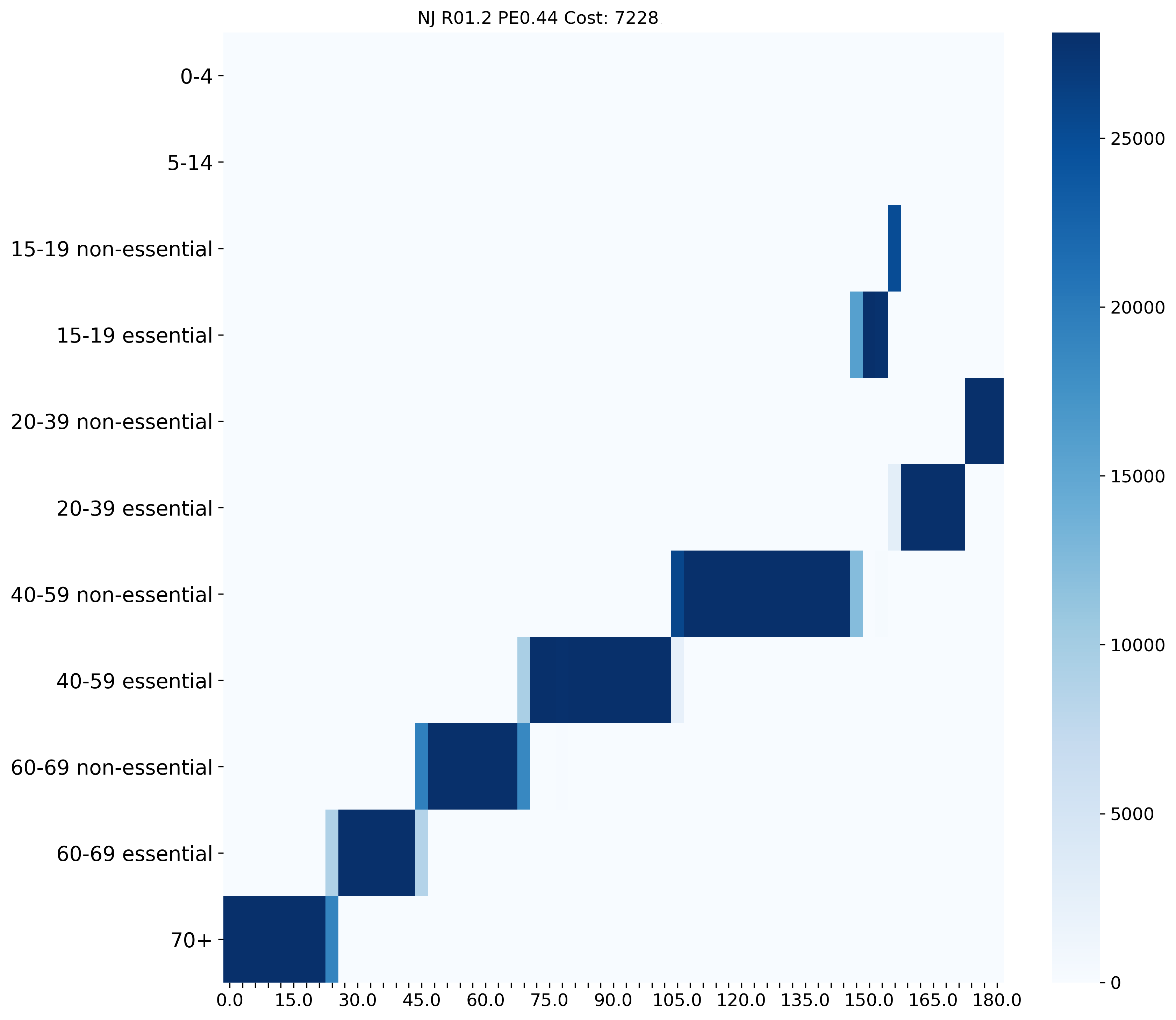}
\caption{Optimal vaccination strategy for Reproduction number 1.2, Percent of workers considered essential 44.} \label{fig:heatmap}
\end{center}
\end{figure}
\begin{figure}
\begin{center}
\includegraphics[angle=0,
width=14.0cm, height=14cm,
trim={0cm 0cm 0cm 0cm},clip]{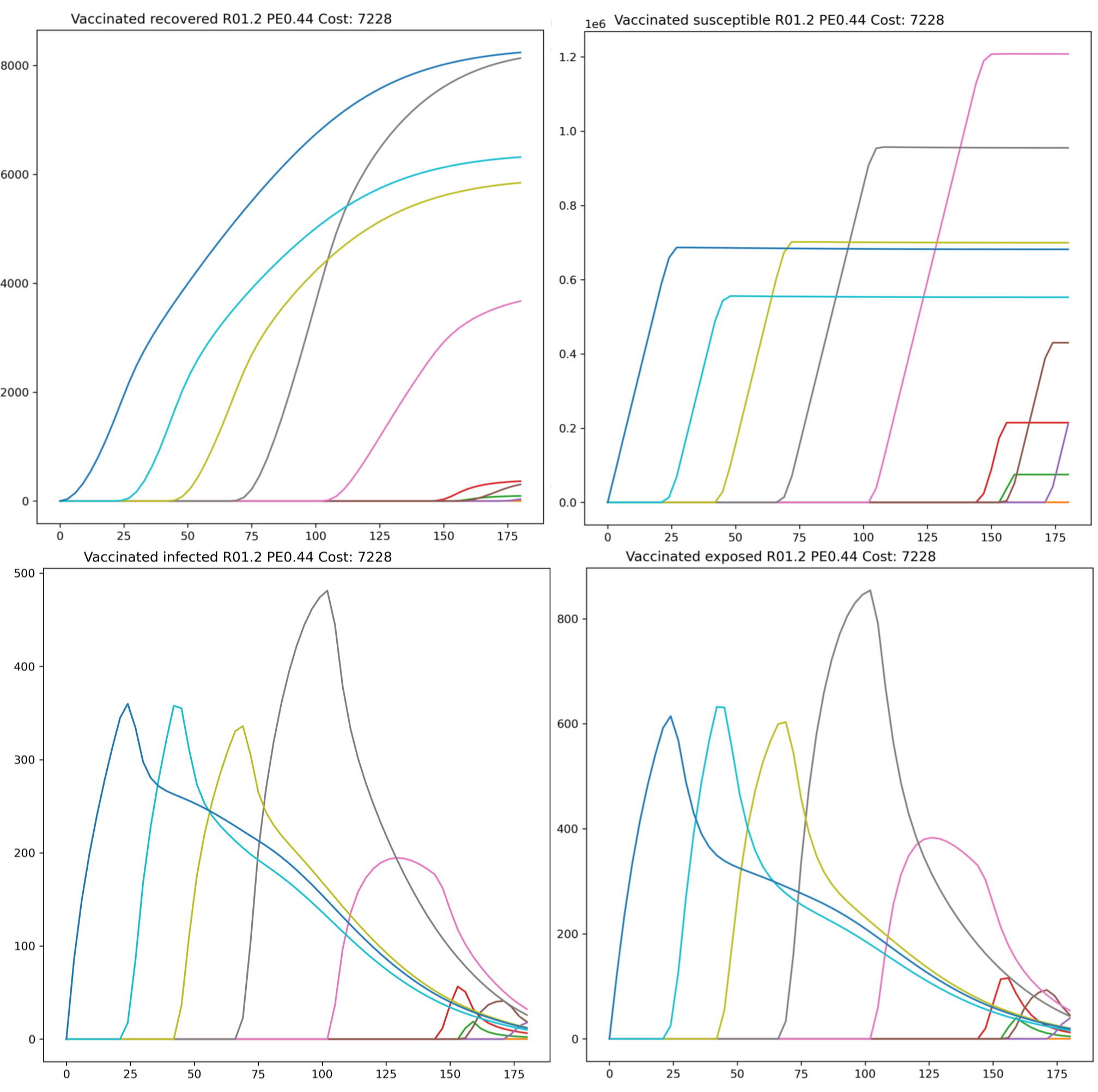}
\caption{\RWC{Population dynamics of the vaccinated compartments: Susceptible, Vaccinated, Exposed vaccinated, Infected vaccinated, and Recovered vaccinated. }} \label{fig:VaccPop}
\end{center}
\end{figure}

While this tool can be fit to any data given information about parameters, we have chosen state data in the United States to test our model. See in Figure \ref{fig:heatmap} an example of an optimal strategy for New Jersey. in Figures \ref{fig:pops} and \ref{fig:VaccPop} are also the accompanying plots of the S, E, I, R, Sv, Ev, Iv and Rv populations. We see from the figures that the compartments act as expected. Notice in the infected population in Figure \ref{fig:pops} each infected population grows at the beginning of the time span, but rapidly approaches zero as it is being vaccinated. \RWC{This is because once somebody is a member of this compartment, s/he can either move to removed (by recovering from the disease or passing away) or to vaccinated infected. 
The latter may occur since the model assumes that an infected person is able to become vaccinated, an event very possible for asymptomatic carriers. As the compartment empties, new infected do appear from the susceptible population, however, as we rapidly vaccinate the susceptible populations, the infected diminish as they have no source of vulnerable people.}

This strategy seems robust not only amongst different values of beta, distributions of essential workers and chosen $R0$, but also amongst different states. It seems as though for the COVID–19 pandemic, the optimal vaccination policy was the one chosen by most countries: vaccinate the oldest population first and work down based on age. In total we have 27 simulations from each of the states tested. In every run, the optimizer chose to vaccinate the oldest population first.

The goal of this paper is to demonstrate how a compartmental epidemiological model can be used in an optimal control problem to minimize casualties when vaccinating the general population. The main limitation of the results is the less realistic assumption that a municipality be able to vaccinate an entire sub-population with ease. See in the susceptible population in Figure \ref{fig:pops} how the susceptible population approaches zero because all people are either vaccinated or have been infected. The reality is that breaking a threshold of something closer to $70\%$ of a population being fully vaccinated would be an optimistic goal for any age group. In fact, we see live in the United States a slowing of vaccination despite the fact that we seemingly have not yet achieved herd immunity.

Note that the cost here is the number of deaths which the population suffers in our time range of 180 days starting vaccination at day one. In Figure \ref{fig:NJTests} there is a vaccination schedule for the same parameters as the above plots, with essential workers of the largest working population being vaccinated first, a policy that many may believe to be optimal. Notice that the number of deaths is larger if one were to use this schedule rather than the optimal one. This is noteworthy as some countries such as Indonesia chose to prioritize their essential workers over their elderly.

Despite these shortcomings, this program could be updated and adjusted to fit a multitude of situations such as these simply by adjusting the equations of the model. When the COVID-19 vaccine arrived, many countries chose this optimal solution of vaccinated on an age-based schedule while others chose to vaccinate their essential workers first. Programs such as ours set the framework to make a more informed decision in future times of crisis. 

All code used in the design and implementation of our epidemiological model is placed on GitHub for free use, note that all programs are written in the Python programming language.

\section{Acknowledgements}
The authors
acknowledge the support of the NSF CMMI project 
\# 2033580 "Managing pandemic by managing mobility".
R.W., S.T.M. and B.P. acknowledge
the support of the Joseph and Loretta Lopez Chair endowment.

\bibliography{scifile}
\bibliographystyle{plain}

\end{document}